\documentclass[%
 reprint,
nofootinbib,
 amsmath,amssymb,
 aps,
]{revtex4-2}

\usepackage{graphicx}
\usepackage{dcolumn}
\usepackage{bm}

\usepackage{braket} 
\usepackage{color}
\usepackage{amsmath, amssymb}

\newcommand{\del}{\partial}
\newcommand{\ii}{{\rm i}}
\newcommand{\dd}{{\rm d}}
\newcommand{\norm}[1]{\left\lVert #1 \right\rVert}
\newcommand{\abs}[1]{\left\lvert #1 \right\rvert}

\DeclareMathOperator{\real}{Re}
\DeclareMathOperator{\imag}{Im}

\begin{document}

\title{Quench dynamics of the Schwinger model via variational quantum algorithms}

\author{Lento Nagano}%
\email[]{lento(at)icepp.s.u-tokyo.ac.jp}
\affiliation{%
International Center for Elementary Particle Physics (ICEPP), The University of Tokyo, 7-3-1 Hongo, Bunkyo-ku, Tokyo 113-0033, Japan
}%
\author{Aniruddha Bapat}%
\affiliation{%
Lawrence Berkeley National Laboratory, 1 Cyclotron Road, Berkeley 94720, CA}%
\author{Christian W Bauer}%
\email[]{cwbauer(at)lbl.gov}
\affiliation{%
Lawrence Berkeley National Laboratory, 1 Cyclotron Road, Berkeley 94720, CA}%

\date{\today}

\begin{abstract}
We investigate the real-time dynamics of the $(1+1)$-dimensional U(1) gauge theory known as the Schwinger model via variational quantum algorithms.
Specifically, we simulate quench dynamics in the presence of an external electric field.
First, we use a variational quantum eigensolver to obtain the ground state of the system in the absence of an external field. With this as the initial state, we perform real-time evolution under an external field via a fixed-depth, parameterized circuit whose parameters are updated using McLachlan's variational principle. 
We use the same Ansatz for initial state preparation and time evolution, by which we are able to reduce the overall circuit depth.
We test our method with a classical simulator and confirm that the results agree well with exact diagonalization.
\end{abstract}

\maketitle

\section{Introduction}
Lattice gauge theory is a powerful tool for studying quantum field theory. 
In the conventional approach, simulations are performed using the Monte-Carlo method, which requires the exponential of the action $\exp(\ii S)$ to be positive and real. This protocol suffers from a sign problem when we consider e.g. a topological term, a finite chemical potential, and real-time dynamics. 

Instead of Monte Carlo, one can use the Hamiltonian formalism, which avoids the sign problem as it is not a sampling-based approach.
However, the quantum state grows exponentially with the size of the system, making this approach not feasible on classical computers.
The advantage of quantum computers is that the computational resources can be kept logarithmic in system size, as was shown in 
the seminal paper by Jordan, Lee, and Preskill~\cite{Jordan:2011ne}. Since that work, digital quantum simulation in the context of quantum field theory has been attracting a lot of interest~\cite{
Jordan:2011ci,Jordan:2014tma,
Martinez:2016yna,Muschik:2016tws,Klco:2018kyo,Kokail:2018eiw,Magnifico:2019kyj,Chakraborty:2020uhf,PhysRevD.105.014504, 2022PTEP.2022c3B01H,2021arXiv210608394D,Yamamoto:2021vxp, 2020PhRvR...2b3342K,Gustafson:2019vsd, 2021PhRvD.103e4507G, 2020Quant...4..306S,PhysRevResearch.4.023176,PhysRevD.106.054509,PRXQuantum.3.020324,2022arXiv220508860T,
PhysRevD.105.074504,2021arXiv210208920A,2022arXiv220703473A,Mezzacapo:2015bra,
Marcos:2014lda,Klco:2019evd,2021PhRvD.103i4501C,PhysRevD.106.114511,2022arXiv221110497K, 
PRXQuantum.3.020320,2021PRXQ....2c0334P,
Klco:2018zqz, 2021PhRvA.103d2410B, 
Garcia-Alvarez:2014uda,Wiese:2014rla,Gustafson:2019mpk,2021arXiv210712769K,2022arXiv220600685P,2022arXiv221105607G,2021PhRvD.104a4512E,2020arXiv201209194S,2020JHEP...12..011L,2020arXiv201106576B,2020PhRvA.102e2422K,Klco:2019xro,PhysRevD.106.114515,2022arXiv221214030D,2021PhRvD.104g4505D,
2022arXiv220813112D,2021arXiv210511548S,2022arXiv221114550Y, 2021PhRvL.127u2001B,2021PhRvD.104h6013L,2021JHEP...07..140G,2020PhRvL.125p0503M,2020arXiv201007965A,Lamm:2019uyc,Mueller:2019qqj,2020arXiv200615746B,Lamm:2018siq,Alexandru:2019ozf,Macridin:2018gdw}.
In particular, real-time simulation is one of the important applications since it can in general not be captured efficiently by any known classical method.
The standard simulation method on quantum computers uses Suzuki-Trotter decomposition, where the circuit depth increases as the evolution time does, which causes a decoherence problem on noisy intermediate-scale quantum (NISQ) devices. Efficient algorithms have been proposed for the preparation of the ground state of certain classes of quantum systems, such as Quantum Imaginary Time evolution and a Quantum Lanczos algorithm~\cite{motta2020determining}. Variational algorithms combine quantum computations with classical optimizations, and are able to perform both state preparation and time evolution using an approach called variational quantum simulation (VQS), even if the accuracy of the method depends on the chosen variational Ansatz. 
A VQS method based on Mclachlan's variational  principle (MVP) was proposed in~\cite{li2017efficient,yuan2019theory}
in which the evolved states are approximated by parameterized states (Ansatz) with a fixed depth\footnote{See~\cite{PhysRevLett.125.010501,PhysRevResearch.2.033281,
PRXQuantum.2.030307,
2020npjQI...6...82C,2022npjQI...8..135G,2020arXiv200902559C,
2021PhRvR...3c3083B,
PRXQuantum.2.010342,2021npjQI...7...79B,
2021arXiv210104579B,2019arXiv190408566H,2022PhRvR...4b3072S,PhysRevA.104.042418,2021arXiv210107677L,2022ScPP...12..122L,
2022PhRvR...4d3161B,PhysRevA.105.062421} for algorithmic developments
of the original algorithm
and
other variational methods for real-time simulation.
See also~\cite{Nat.Comput.Sci.3.25.2023} for a recent review.}. 

\begin{figure}
    \centering
    \includegraphics[width=80mm]{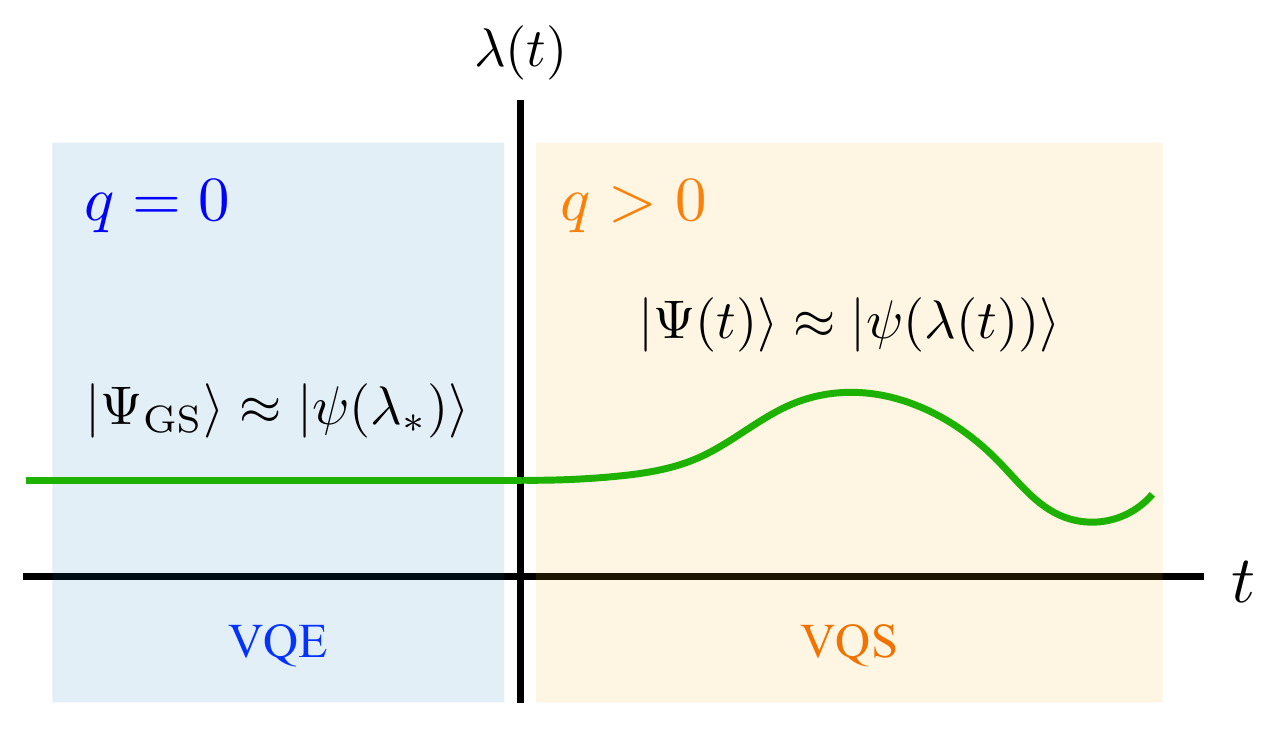}
    \caption{Sketch of our simulation. We start from the ground state in the absence of external electric field~$q$. We then suddenly turn on the external field~$q$ and evolve the state via the Hamiltonian with $q>0$. These states are approximated by the same Ansatz $\ket{\psi(\lambda)}$.}
    \label{fig:protocol}
\end{figure}

In this work, we apply this variational method to investigate the real-time dynamics of $(1+1)$ dimensional U(1) gauge theory called the Schwinger model\footnote{The authors~\cite{2022MLS&T...3d5030L} proposed an application of the variational method to a scalar field theory and performed adiabatic state preparation, while they did not provide an explicit real-time simulation.
}.
Specifically, we perform real-time simulation after turning on an external electric field to see electron-positron pair creations induced by the external field, which is the so-called Schwinger mechanism~\cite{PhysRev.82.664}.
This is similar to what was considered in~\cite{buyens2017real} where a classical tensor network approach was used~\footnote{See also~\cite{PhysRevD.106.116007} for a recent study in a slightly different setup by using VQE and Suzuki-Trotter decomposition.}.
See Fig.~\ref{fig:protocol} for the sketch of our simulation protocol. 
We first prepare the ground state~$\ket{\Psi_\text{GS}}$ in the absence of the external electric field~$q$ by using a variational quantum eigensolver (VQE).
At initial time~$t=0$ the external field is then suddenly turned on (the quantum quench), and the time evolution in the system with an external electric field is then studied.
We approximate the dynamical states~$\ket{\Psi(t)}=e^{-\ii H_{q\neq0}t}\ket{\Psi_\text{GS}}$ by the same Ansatz used in VQE, and evolve the parameters according to MVP.
Note that by performing both state preparation and time evolution through variational circuits, the depth of a quantum circuit is greatly reduced.

The structure of this paper is organized as follows.
Section~\ref{sec:Schwinger-model} introduces the Hamiltonian of the Schwinger model and observables we will focus on.
Section~\ref{sec:method} explains a method we will use for the simulation. In Section~\ref{sec:results} results are presented and compared them with exact diagonalization.
Finally, conclusions are given in Section~\ref{sec:summary}.

\section{The Schwinger model}
\label{sec:Schwinger-model}

The Schwinger model describes quantum electromagnetism in one spatial and one time dimension. This model is relatively simple, and can in fact be solved analytically~\cite{Schwinger:1962tp, LOWENSTEIN1971172} in the massless limit. It is nevertheless a very interesting field theory to study, since despite its simplicity it shares several features with the QCD, the theory of the strong interaction, such as confinement and charge screening~\cite{Coleman:1975pw,Gross:1995bp}.
\subsection{Lattice Hamiltonian and spin description}
Here we will define the lattice Hamiltonian and introduce its spin description.
We mostly follow the convention used in~\cite{PhysRevD.105.014504}. First of all,
the Lagrangian of the continuum Schwinger model is 
given by 
\begin{align}
\mathcal{L}_{\rm con}
&= -\frac{1}{4} F_{\mu\nu} F^{\mu\nu} +\ii\bar{\psi}\gamma^\mu (\del_\mu +\ii gA_\mu -m) \psi
\nonumber\\
&\qquad
+\frac{g\theta}{4\pi} \epsilon_{\mu\nu} F^{\mu\nu}
\,.
\end{align}
Here the first two terms correspond to the kinetic term of the gauge boson and fermion, respectively, while the third term denotes a topological term that does not affect the classical equations of motion, but does affect the quantum spectrum. 

Taking the gauge $A_0=0$ and introducing the canonical momentum $\Pi:=\del \mathcal{L}_{\text{con}}/\del(\del_0 A^{1})$, we can write the continuum Hamiltonian as 
\begin{align}
\int dx \left[ \frac{1}{2} \Big( \Pi -\frac{g\theta}{2\pi} \right)^2
-\ii \bar{\psi} \gamma^1 (\partial_1 +\ii g A_1 - m) \psi 
\Big]
\,,
\nonumber
\end{align}
with $\Pi=\partial_0 A^1 +g\theta/2\pi$. As is usual in $A_0=0$ gauge, Gauss's law has to be enforced through an extra constraint, and
physical states have to satisfy~$G\ket{\text{phys}}=0$ with $G=\partial_1 \Pi +g\psi^\dag \psi$.

A lattice version of this Hamiltonian can be obtained following the work of~\cite{Kogut:1974ag}.  
Fermions are put on a staggered lattice, where the position $x$ is sampled at discrete points $x_n$. Here $n=0,\ldots,N-1$ label the lattice sites corresponding to $x_n=na$, and $a$ is the lattice spacing.
The fermion fields at each lattice site are written in terms of $\chi_n$, which represents the Dirac fermion~$\psi(x)=\big(\psi_u(x),\psi_d(x) \big){}^\mathsf{T}$ through 
\begin{align}
\frac{\chi_n}{\sqrt{a}} \ \longleftrightarrow \ \left\{ 
\begin{array}{ll} 
\psi_u (x_n) & (n:{\rm even}) \cr
\psi_d (x_n) & (n:{\rm odd}) \end{array}
\right.
\,. 
\end{align}
The gauge fields are represented through operators living on the links between $n$-th and $(n+1)$-th lattice sites 
\begin{align}
   U_n &  \longleftrightarrow e^{-\ii ag A^1 (x_n)}\,,
   \\
   L_n & \longleftrightarrow -\Pi(x_n)/g\,.
\end{align}
These lattice variables satisfy the  commutation relations
\begin{align}
\{ \chi_n^\dag ,\chi_m \} &= \delta_{mn}\,,\nonumber\\
\{ \chi_n ,\chi_m \} &=0\,,\nonumber\\   
[U_n ,L_m] &=\delta_{mn}U_n\,,
\nonumber
\end{align}
and $U_n^\dag=U_n^{-1}, L_n^\dag=L_n$.
With these definitions, the lattice Hamiltonian is given by
\begin{align}
&H 
= J \sum_{n=0}^{N-2} \left( L_n +q  \right)^2 
 -\ii w \sum_{n=0}^{N-2} \big( \chi_n^\dag U_n \chi_{n+1} 
 \nonumber
\\
&\qquad\qquad
 -\chi_{n+1}^\dag U_n^\dag \chi_{n} \big)  +m\sum_{n=0}^{N-1} (-1)^n \chi_n^\dag \chi_n
 \,,
 \label{eq:fermion-Hamiltonian}
\end{align}
where $w=1/(2a)$, $J=g^2a/2$ and $q=\theta/(2\pi)$.
Introducing nonzero $q$ corresponds to turning on the external electric field.

Gauss's law gives a constraint for the~$L_n$ links on the lattice given by
\begin{align}
L_n -L_{n-1} =  \chi_n^\dag \chi_n -\frac{1-(-1)^n}{2}\,.
\label{eq:Gauss_lattice}
\end{align}
We impose the open boundary condition $L_{-1}=0$ and fix the gauge $U_n=1$ to eliminate gauge fields from the Hamiltonian.

One can transform the above Hamiltonian into a spin Hamiltonian through the Jordan-Wigner transformation~\cite{Jordan1928},
\begin{equation}
 \chi_n = \frac{X_n-\ii Y_n}{2}\prod_{i=0}^{n-1}(-\ii Z_i)\,.
\end{equation}
This leads to the spin Hamiltonian given by
\begin{align}
 H &= J\sum_{n=0}^{N-2} \left[\sum_{i=0}^{n}\frac{Z_i + (-1)^i}{2}+q\right]^2 
\nonumber
\\
&
 + \frac{w}{2}\sum_{n=0}^{N-2}\big[X_n X_{n+1}+Y_{n}Y_{n+1}\big]
 + \frac{m}{2}\sum_{n=0}^{N-1}(-1)^n Z_n\,,
\label{eq:spin-hamiltonian}
\end{align}
up to an irrelevant constant.

\subsection{Observables}
\label{sec:observables}
While there are several observables one can study in the Schwinger model, in this work we focus on three observables~\cite{buyens2017real,funcke2020topological}.
The first one is the total electric field,
\begin{align}
\mathcal{E}(t)
&=\frac{g}{N}\sum_{n=0}^{N-1}
\Braket{L_{n}+q}_{t}\,,
\end{align}
where $\braket{\bullet}_{t}:=\braket{\psi(t)|\bullet|\psi(t)}$.
In the spin description this is given by
\begin{align}
\mathcal{E}(t)
=\frac{g}{2N}
\sum_{n=0}^{N-1}
\sum_{k=0}^{n}
\braket{Z_{k}}_{t}
+\frac{g}{2N}
\sum_{n=0}^{N-1}\sum_{k=0}^{n}(-)^{k}
+gq\,.
\end{align}

The second observable is the chiral condensate $\braket{\bar{\psi}\psi}$ whose lattice counterpart is given by
\begin{align}
\Sigma(t)
&=
\frac{ag}{N}
\sum_{n=0}^{N-1} (-1)^n \Braket{\chi_n^\dag \chi_n}_{t}
\\&=
\frac{ag}{N}
\sum_{n=0}^{N-1}(-)^{n}\braket{Z_{n}}_{t}\,,
\end{align}
up to an irrelevant constant.
In the heavy mass regime $m\gg g$ this can be interpreted as the expectation value of the particle number operator, while this interpretation is not exact in other regimes.
Nonetheless, this gives an approximate metric for particle-antiparticle creation.

The third and final observable is the U(1) charge $Q$ defined by 
\begin{align}
    Q = \frac{1}{N}\sum_{n=0}^{N-1}\braket{Z_n}_t
    \,.
\end{align}
This observable is useful since it  has to be preserved in the evolution under the Hamiltonian~\eqref{eq:spin-hamiltonian}.

\section{Method}
\label{sec:method}
\subsection{Ansatz} 
As already discussed, this study uses variational quantum circuits for both the state preparation and the time evolution of the system after the quantum quench.
To create the initial ground state in the theory without an external electric field
we use the  Hamiltonian Variational Ansatz (HVA)~\cite{PhysRevA.92.042303, 10.21468/SciPostPhys.6.3.029, PRXQuantum.1.020319} defined as
\begin{equation}
\ket{\psi(\bm{\alpha},\bm{\beta},\bm{\gamma})}
=U_{L-1}\cdots U_{0}V_{\text{init}}\ket{0}\,,
\end{equation}
where 
\begin{align}
V_{\text{init}} &= \prod_{n:\text{even}} X_{n}\,,
\\
U_{l}(\bm{\alpha}_{l},\bm{\beta}_{l},\bm{\gamma}_{l})
&=\prod_{n=0}^{N-1} u^{(Z)}_{n}(\gamma_{l,n})
\notag\\&\quad\times
\prod_{n:\text{odd}}u^{(ZZ)}_{n}(\beta_{l,n})
\prod_{n:\text{even}}u^{(ZZ)}_{n}(\beta_{l,n})
\notag\\&\quad\times
\prod_{n:\text{odd}}u^{(XY)}_{n}(\alpha_{l,n})
\prod_{n:\text{even}}u^{(XY)}_{n}(\alpha_{l,n})\,,
\end{align}
with 
\begin{align}
u^{(Z)}_{n}(\gamma_{l,n})
&=\exp\left[
\ii\frac{\gamma_{l,n}}{2}Z_{n}
\right]\,,\\
u^{(ZZ)}_{n}(\beta_{l,n})
&=\exp\left[
\ii\frac{\beta_{l,n}}{2}Z_{n}Z_{n+1}
\right]\,,\\
u^{(XY)}_{n}(\alpha_{l,n})
&=\exp\left[
\ii\frac{\alpha_{l,n}}{2}\frac{X_{n}X_{n+1}+Y_{n}Y_{n+1}}{2}
\right]\,.
\end{align}
Note that this Ansatz preserves the global U(1) symmetry, which must be preserved for true evolution under the Hamiltonian. Note that in the following discussion the whole set of parameters is often denoted by $\bm{\lambda}$
\begin{equation}
\bm{\lambda}=(\bm{\alpha}_{0}, \bm{\beta}_{0}, \bm{\gamma}_{0}, \cdots ,\bm{\alpha}_{L-1}, \bm{\beta}_{L-1}, \bm{\gamma}_{L-1})\,,
\nonumber
\end{equation}
and a general $u\in\{  u_{n}^{(XY)},u_{n}^{(ZZ)}, u_{n}^{(Z)}\}_{n=0}^{N-1}$ with dependence on these parameters is denoted by $u(\lambda)$.

\subsection{McLachlan's variational principle}
McLachlan's variational principle~\cite{1964MolPh...8...39M} gives the following set of equations
\begin{equation}
\sum_{i,j}M_{ij}\dot{\lambda}_{j} = V_{j}\,,
\label{eq:McLachlan-parameter-evolution-eq}
\end{equation}
where
\begin{align}
M_{ij} &= 2\real\left[ A_{ij}\right] + 
2C_{i}^{(0)}C_{j}^{(0)}
\label{eq:def-Mij}
\,,
\\
V_{i} &= 2\imag\left[ C_{i}\right]
+2\ii C_{i}^{(0)}\braket{H}_{\psi}
\label{eq:def-Vi}
 \,,
\end{align}
with
\begin{align}
\label{McLachlanParameterDefA}
A_{ij} &= \frac{\del\bra{\psi}}{\del\lambda_{i}}\frac{\del\ket{\psi}}{\del\lambda_{j}}\,,
\quad
C_{i} = \frac{\del\bra{\psi}}{\del\lambda_{i}}H\ket{\psi}\,,
\\
C_{i}^{(0)}&=\frac{\del\bra{\psi}}{\del\lambda_{i}}\ket{\psi}
\,,\quad
\braket{H}_{\psi} = \braket{\psi|H|\psi}\,.
\label{McLachlanParameterDefB}
\end{align}

Each term is evaluated on a quantum computer as follows~\cite{yuan2019theory,yuan2019theory}. We use the same variational Ansatz used to create the ground state in the absence of the background electric field for the state after the background electric field is turned on. This allows us to obtain the corresponding Ansatz for the derivatives of the state with respect to the parameters $\lambda$ needed in Eqs.~\eqref{McLachlanParameterDefA} and~\eqref{McLachlanParameterDefB}. 
From the explicit forms of the functions $u(\lambda)$ one obtains
\begin{equation}
\frac{\dd u}{\dd \lambda_{i}}
=
f_{i} u\bm{\sigma}_{i}\,
\label{eq:derivative-ansatz-general}
\end{equation}
where $\bm{\sigma}_{i}\in\{I,X,Y,Z\}^{\otimes N}$ and $f_i$ are complex scalar ``structure constants".
Explicitly the derivatives are given by
\begin{align}
\frac{\dd u_{n}^{(XY)}}{\dd \alpha_{l,n}}
&=\frac{\ii}{2}\cdot u_{n}^{(XY)}
\frac{X_{n}X_{n+1}+Y_{n}Y_{n+1}}{2}\,,
\\
\frac{\dd u_{n}^{(ZZ)}}{\dd \beta_{l,n}}
&=\frac{\ii}{2}\cdot u_{n}^{(ZZ)}Z_{n}Z_{n+1}\,,
\\
\frac{\dd u_{n}^{(Z)}}{\dd \gamma_{l,n}}
&=\frac{\ii}{2}\cdot u_{n}^{(Z)}Z_{n}\,.
\end{align}

Using this information, one finds
\begin{equation}
\frac{\del\ket{\psi}}{\del\lambda_{i}}
=
f_{i}\hat{U}_{i}(\lambda)V_{\text{init}}\ket{0}\, ,
\end{equation}
where $\hat{U}_{i}(\lambda)$ is given by replacing a unitary block $u$ corresponding to $\lambda_i$ in the Ansatz to $(u\bm{\sigma}_{i})$ and $f_{i}$ is defined in~\eqref{eq:derivative-ansatz-general}.
Similarly, the coefficients $M_{ij}$ and $V_{i}$ given in Eq.~\eqref{eq:def-Mij} and Eq.~\eqref{eq:def-Vi} can be evaluated as  
\begin{align}
M_{ij}&=\frac{1}{2}\real\left[\braket{0|\hat{U}^{\dag}_{i}\hat{U}_{j}|0}\right]
\notag\\&\qquad
-\frac{1}{2}\real\left[\braket{0|\hat{U}^{\dag}_{i}U|0}\right]\real\left[\braket{0|\hat{U}^{\dag}_{j}U|0}\right]\,,
\\
V_{i} &= -\sum_{p}h_{p}\real\left[\braket{0|\hat{U}^{\dag}_{i}\bm{\sigma}_{p}U|0}\right]
\notag\\&\qquad
+\real\left[\braket{0|\hat{U}^{\dag}_{i}U|0}\right]
\braket{H}_{\psi}
\,.
\end{align}
Note that the Hamiltonian can be decomposed into Pauli strings as $H=\sum_{p}h_{p}\bm{\sigma}_{p}$.
Each term in the above equations is therefore evaluated by the quantum circuit given in Fig.~1 of~\cite{yuan2019theory}.
The initial state in the ancilla qubit is $(\ket{0}+\ket{1})/\sqrt{2}$ corresponding to $\theta=0$.

Some more details on the McLachlan variational principle are given in App.~\ref{app:McLalchan}.

\subsection{Quench dynamics via VQE and VQS}
This section summarizes again the steps required to simulate quench dynamics using VQE and VQS variational algorithms. 
One starts from the ground state in the absence of the external electric field $\ket{\Psi_{\text{GS}}(q=0)}$.
One then turns on the external field $q\neq0$ and trace the time evolution, $\ket{\Psi(t)}=e^{-\ii H_{q\neq0}t}\ket{\Psi_{\text{GS}}(q=0)}$.

This process is implemented through the following quantum variational protocol.
\begin{enumerate}
\item
State preparation via VQE:
One approximates $\ket{\Psi_{\text{GS}}(q=0)}$ by $\ket{\psi(\lambda_{\text{opt}})}$, and determines $\lambda_{\text{opt}}$ by minimizing $\braket{\psi(\lambda)|H|\psi(\lambda)}$ on a classical computer.
\item
Real-time evolution via McLachlan's variational principle:
One uses $\lambda_{\text{opt}}$ as initial values and evolve $\lambda$ via~\eqref{eq:McLachlan-parameter-evolution-eq}.
The coefficients $M_{ij}$ and $V_i$ are evaluated by a quantum circuit while the parameter evolution is done by a classical computer.
\end{enumerate} 
In a standard algorithm, one need both a state preparation and time-evolution circuit, but using the approach presented here one reduces the depth by using the same Ansatz for both processes. 
\section{Results}
\label{sec:results}
This section presents our results of the simulation using the variational algorithms and compares them against results obtained from exact diagonalization (ED).
The VQE and VQS results are obtained from noiseless state-vector simulation implemented by Qulacs~\cite{2021Quant...5..559S}, while ED results are obtained by QuSpin~\cite{2017ScPP....2....3W}.

\subsection{Ground state preparation via VQE}
\begin{figure}[htbp]
    \centering
    \includegraphics[width=85mm]{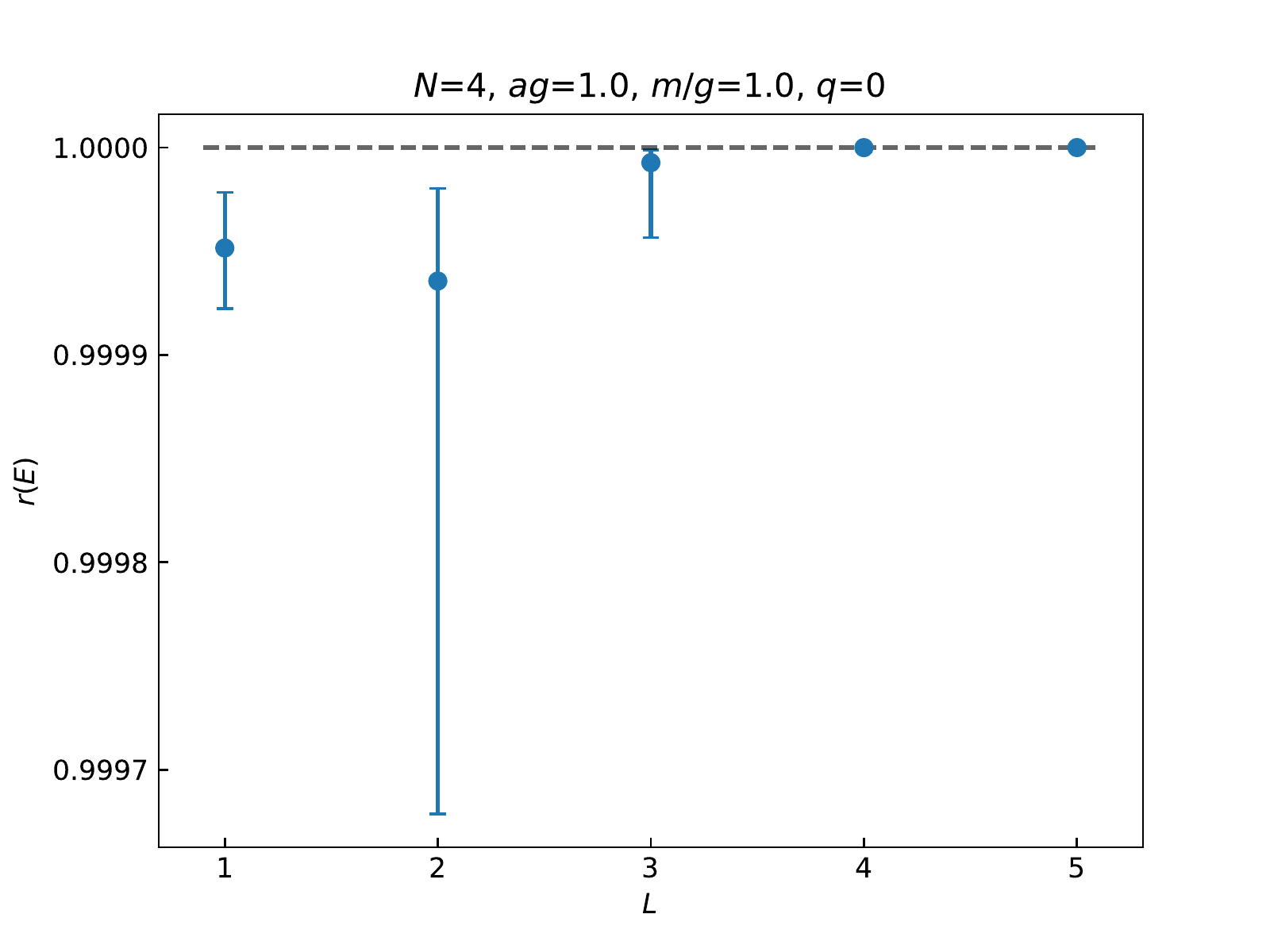}
    \caption{Ground state preparation via VQE: a metric of accuracy $r(E):=(E_\text{max}-E_{\text{VQE}})/(E_\text{max}-E_{\text{min}})$. Dots/error bars show the median and 25-75 percentiles of 20 samples.}
    \label{fig:vqe-plot}
\end{figure}

We first perform VQE for $N=4$, $ag=1$, $m/g=1$ in the absence of the external field $q=0$. 
We repeat optimizations $20$ times starting from different random initializations. 
Fig.~\ref{fig:vqe-plot} shows a metric of accuracy~\cite{2020PNAS..11725396P}~$ r(E):=(E_\text{max}-E_{\text{VQE}})/(E_\text{max}-E_{\text{min}})$ as a function of the number of layers, where $E_\text{max/min}$ is the highest/lowest eigenvalue of the Hamiltonian~\footnote{This ratio takes $0$ for the worst case ($E_{\text{VQE}}=E_\text{max}$) and $1$ for the best case ($E_{\text{VQE}}=E_\text{min}$).
We obtain $E_\text{max/min}$ via ED.
}. The central value corresponds to the median of the 20 optimizations performed, while the error bar represents the 25-75 percentiles. One observes that high accuracy $r(E)\geq 0.999$ can be achieved for all $L$ and that for $L\geq 4$ the uncertainties improve markedly. 

\subsection{Quench dynamics via VQS}
After preparing the initial state, we perform VQS for~$N=4$, $ag=1$, $m/g=1$, and $q=2$~\footnote{We regularize the matrix $M_{ij}$ as $M\to M+\epsilon I$ if $\det (M)<\epsilon$ when we perform a matrix inversion. In the following simulation, we set $\epsilon=10^{-7}$.}.
First, we investigate the dependence of systematic errors on the number of layers $L$ and a time increment $\delta t$.
\begin{figure*}[htbp]
\centering
\includegraphics[width=160mm]{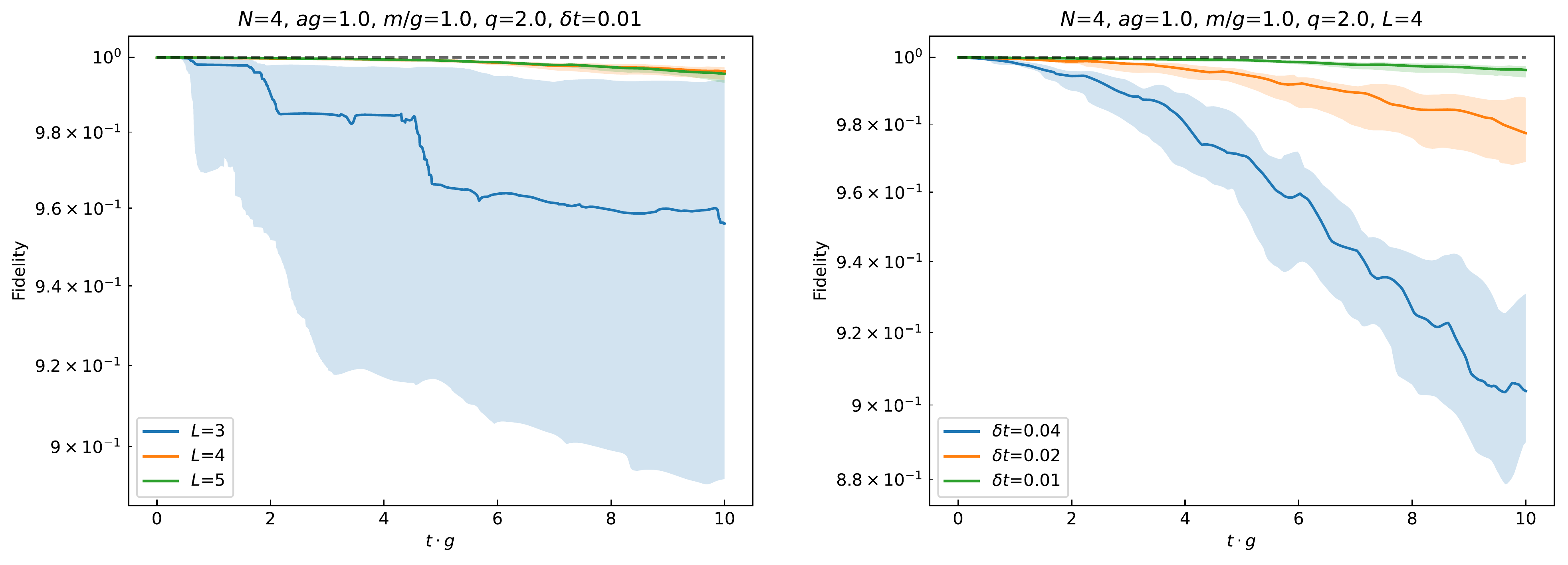}
\caption{Fidelity between states from VQS and ED for $N=4, a\cdot g=1.0, m/g=1.0, q=2.0$. Solid curves/error bands show the medians and 25-75 percentiles of 20 samples: (left) dependence of the number of layers $L\in\{3,4,5\}$ with $\delta t =0.01$ fixed, (right) dependence of a time increment $\delta t\in\{0.01,0.02,0.04\}$ with $L$ fixed.}
\label{fig:fidelity-L-and-dt-deps}
\end{figure*}
The left plot in Fig.~\ref{fig:fidelity-L-and-dt-deps} shows the fidelity $F(t):=\abs{\braket{\Psi_{\text{ED}}(t)|\psi_{\text{VQS}}(t)}}^2$ between the states obtained from VQS and ED as a function of (coupling constant times) time for different number of layers.
One can see again that the uncertainty improves dramatically as the number of layers is raised above 3 and that for $L \geq 4$ the (median) fidelity is above $0.99$
and is improved by increasing the number of layers.
The right panel shows the same plot, but this time varying~$\delta t$ for fixed $L$. We see that the VQS results can be improved significantly by decreasing $\delta t$.

\begin{figure*}[htbp]
\centering
\includegraphics[width=170mm]{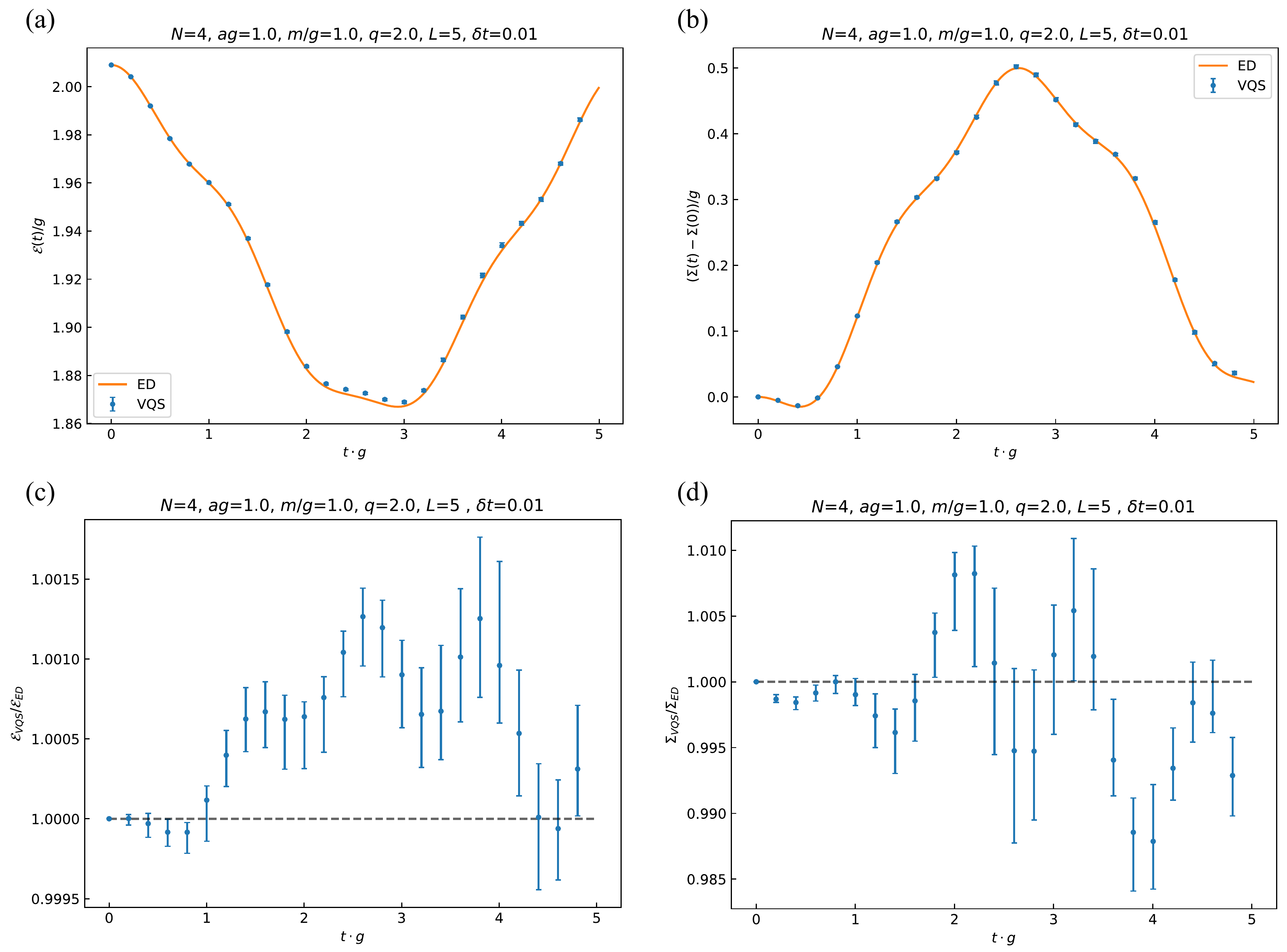}
\caption{
Dynamics of physical observables for $N=4, a\cdot g=1.0, m/g=1.0, q=2.0$ with $L=5$ and $\delta=0.01$. Dots/error bars show the median and 25-75 percentiles of 20 samples: (a) electric field, (b) chiral condensate, (c), (d) ratio between the values of observables obtained from ED and VQS}
\label{fig:obs-vqs}
\end{figure*}

Next, we evaluate the physical observables discussed in Section~\ref{sec:observables} and compare them to the results obtained by exact diagonalization. We verified that the U(1) charge agrees perfectly with the exact result, as can be expected since our Ansatz satisfies the global U(1) symmetry of the problem. The remaining two observables are shown in Fig.~\ref{fig:obs-vqs} with $L=5$ and $\delta t =0.01$ fixed.
The VQS results are consistent with those from ED up to a few $\%$ errors. 
The errors from the variation over the 20 initial conditions is of the same order of magnitude as the difference from the exact result, but for the electric field and $1.5 \lesssim t\cdot g \lesssim 4.5$ the difference between the exact result and the central value of the VQS results are about three times the size of the quoted error. 

For $t\cdot g\lesssim3$, the value of the electric field decreases while that of the chiral condensate increases, followed by the oscillation.
This can be interpreted as follows: the external electric field first provides energy for fermions and then leads to particle pair creations.

\section{Summary and discussion}
\label{sec:summary}
In this work, we demonstrated a possible application of the variational quantum algorithm to a gauge theory.
Specifically, we investigated the real-time dynamics in the Schwinger model after suddenly turning on the external electric field, by combining VQE and VQS methods.
We performed the (classically-emulated) state-vector simulation and found that the results obtained from the quantum algorithms are consistent with those obtained from ED.
Our simulation results can be interpreted as a population of a particle-anti-particle pair induced by the external field.

There are many possible future directions.
This paper used the original algorithm proposed by Li and Benjamin~\cite{li2017efficient}.
There are two main drawbacks to this approach: First, the matrix $M$ can be singular or ill-conditioned in practice, leading to unstable trajectories. Workarounds such as regularization add a parameter that must be tuned. Secondly, computing the each entry of $M$ requires $O(N_p^2)$ calls to the quantum computer
where $N_p$ is the number of parameters. 
There are many attempts to overcome this problem~\cite{
2020npjQI...6...82C,2022npjQI...8..135G,2020arXiv200902559C, 
2021PhRvR...3c3083B, 
PRXQuantum.2.010342,2021npjQI...7...79B,
2021arXiv210104579B,2019arXiv190408566H,2022PhRvR...4b3072S,PhysRevA.104.042418,2021arXiv210107677L,2022ScPP...12..122L,
2022PhRvR...4d3161B}. It would be important to see if these methods can improve our simulation results in terms of accuracy and measurement cost.

Toward an implementation on real quantum devices, it is important to understand the effects of hardware noise and statistical error coming from a finite number of measurements. Besides, combination with error mitigation methods can be an essential ingredient.

Finally, it would be interesting to consider an extension to the higher dimensional and/or non-Abelian gauge theory.
For this purpose, a careful search for an Ansatz that is efficient and preserves gauge invariance during simulation can be crucial. 
\begin{acknowledgements}
LN would like to thank N.~Gomes and M.~Honda for useful conversations.
He would also like to thank H.~Taya for his seminar on the Schwinger mechanism. 
This study is partly carried out under the
project “Optimization of HEP Quantum Algorithms” supported by the U.S.-Japan Science and
Technology Cooperation Program in High Energy
Physics. CWB and AB also acknowledge support from the DOE, Office of Science under contract DE-AC02-05CH11231, through Quantum Information Science Enabled Discovery (QuantISED) for High Energy Physics (KA2401032)
\end{acknowledgements}

\appendix
\section{McLachlan's variational principle}
\label{app:McLalchan}

This variational principle starts from a variational Ansatz state $\ket{\psi(\lambda)}$ given as the output of a parameterized circuit (and a global phase $\lambda_0$),
\begin{equation}
    \ket{\psi(\lambda)} = e^{i\lambda_0(t)}\prod\limits_{i=1}^{N_p}U(\lambda_i(t))\ket{0}
    \,.
\end{equation}
In other words, the time dependence of the state is encoded through the time dependence of the parameters $\lambda(t)$. 
The task of simulating Schr\"odinger evolution $\ket{\Psi(t)} = \mathcal T e^{-i\int H(s)ds}\ket{\Psi(0)}$ under a general, time-dependent Hamiltonian $H(t)$ using the Ansatz state $\ket{\psi(\lambda)}$ reduces to finding the parameter function $\lambda(t)$ such that at any time $t$, the state $\ket{\psi(\lambda(t))}$ optimally approximates the exact state $\ket{\Psi(t)}$. 

A standard approach to this problem is to use a dynamical variational principle such as the McLachlan's variational principle, or MVP for short. Other choices such as the Dirac-Frenkel variational principle and the time-dependent variational principle exist, and while different in subtle ways, they all agree under certain mild assumptions. In this section we focus on the MVP.

The central idea behind MVP is to minimize the difference between the rates of change of $\ket{\psi}$ under exact Hamiltonian evolution and variational evolution due to ${\rm d}\lambda / {\rm d} t$. Normally, this is expressed as minimizing the variation of the norm difference shown below,
\begin{equation}
\label{eq:McLachlanVP}
    \delta\norm{\left(\frac{d}{dt} + iH\right)\ket{\psi(\lambda(t))}}^2 = 0
\end{equation}
where $\norm{\cdot}$ is the $l_2$-norm. Note that the variation in the above is with respect to~$\delta({\rm d}\lambda_i / {\rm d} t)$, i.e., total time derivatives in each parameter $\lambda_i$. In other words, one is looking for stationary points by varying the tangent vector ${\rm d}\lambda / {\rm d} t$.
Since one is dealing with a time-dependent Hamiltonian, the equation has to be written a little more carefully as 
\begin{equation}
    \lim_{\Delta t\rightarrow 0}\delta\frac{\norm{(U(t+\Delta t,t) - I)\ket{\psi}- \Delta t\ket{\dot\psi}}^2}{\Delta t^2}  = 0\, ,
\end{equation}
where $U(t',t):= \mathcal T e^{-i\int_t^{t'} H(s)ds}$ and $I$ is the identity matrix. The main difference from the time-independent case is due to the appearance of an extra term at order $\Delta t^2$ due to the time dependence of $H$,
\begin{align}
    U(t+\Delta t,t) &= I - iH\Delta t - (H^2 + i\dot H)\frac{\Delta t^2}{2} + O(\Delta t^3) \\
    &\equiv U_{TI}(t+\Delta t,t) - i\dot H\frac{\Delta t^2}{2} + O(\Delta t^3)
\end{align}
where $U_{TI}$ is the propagator if $H$ were held constant in time at the value $H(t)$. However, since the limit is insensitive to terms above linear order, the time-dependence of $H$ can be safely ignored. 

Before expanding~\eqref{eq:McLachlanVP}, one can implement some constraints on $\ket{\psi}$ and its time derivatives due to the normalization condition $\braket{\psi\vert\psi} = 1$. Setting the first and second time derivatives to zero yields, respectively,
\begin{align}
    \braket{\dot\psi\vert\psi} &= - \braket{\psi\vert\dot\psi}\\
    \real\braket{\ddot\psi\vert\psi} &= - \braket{\dot\psi\vert\dot\psi}\, .
\end{align}
Now, one expands the norm of the difference vector,
\begin{align}
    \norm{\ket{\dot\psi} + iH\ket{\psi}}^2 &= \left(\bra{\dot\psi} - i\bra{\psi}H\right) \left(\ket{\dot\psi} + iH\ket{\psi}\right)\\
    &= \braket{\dot\psi\vert\dot\psi} - 2\imag\braket{\psi\vert H\vert\dot\psi} + \braket{H^2}
    \,,
\end{align}
using the notation $\braket{O} = \braket{\psi\vert O\vert\psi}$.
Next, one writes $\ket{\dot\psi}$ in terms of partial derivatives
\begin{equation}
  \ket{\dot\psi} = \dot\lambda_0\frac{\partial}{\partial\lambda_0}\ket\psi + \sum\limits_{i=1}^{N_p}\dot\lambda_i\frac{\partial}{\partial\lambda_i}\ket{\psi} = i\dot\lambda_0\ket\psi + \dot\lambda_i\ket{\partial_i\psi}\, , 
\end{equation}
where $\ket{\partial_i\psi} = \frac{\partial}{\partial\lambda_i}\ket\psi$. The last expression is implicitly summed over $i$ from $1$ to $N_p$, and the $i=0$ term (corresponding to global phase) gives a derivative parallel to $\ket\psi$. The purpose of the~$\lambda_0$ term is to keep track of variations parallel to~$\psi$ which do not change the overall state but can have an effect on the dynamics of the  variational parameters. In practice, including it can lead to more well-behaved dynamics. 

Then, one can separately express the terms containing~$\ket{\dot\psi}$ as
\begin{align}
    \braket{\dot\psi\vert\dot\psi} &= \dot\lambda_i\dot\lambda_j\real\braket{\partial_i\psi\vert\partial_j\psi} 
    \notag\\&\quad
    + \dot\lambda_0\dot\lambda_i\imag\braket{\psi\vert\partial_i\psi} + \dot\lambda_0^2\, ,\\
    \imag\braket{\psi\vert H\vert\dot\psi} &= \dot\lambda_0\braket{H} + \dot\lambda_i\imag\braket{\psi\vert H\vert\partial_i\psi}\, ,
\end{align}
where the properties $\braket{\psi\vert\psi}=1$ and $\braket{\psi\vert H\vert\psi}= \braket{H}$ is real have been used. 
Taken together, this yields
\begin{align}
    \norm{\ket{\dot\psi} + iH\ket{\psi}}^2 &= \dot\lambda_i\dot\lambda_j\real\braket{\partial_i\psi\vert\partial_j\psi}
    + \dot\lambda_0\dot\lambda_i\imag\braket{\psi\vert\partial_i\psi}
    \notag\\&\quad
    + \dot\lambda_0^2 
    - 2\dot\lambda_0\braket{H}- 2\dot\lambda_i\imag\braket{\psi\vert H\vert\partial_i\psi}
    \notag\\&\quad
     + \braket{H^2}\, ,\\
    &= \dot\lambda_i\dot\lambda_j\real\braket{\partial_i\psi\vert\partial_j\psi} + \dot\lambda_0\dot\lambda_i\imag\braket{\psi\vert\partial_i\psi}
    \notag\\&\quad
    - 2\dot\lambda_i\imag\braket{\psi\vert H\vert\partial_i\psi} + \left(\dot\lambda_0 - \braket{H}\right)^2
    \notag\\&\quad
    + \sigma^2_{\psi}(H)\, ,
\end{align}
where $\sigma^2_{\psi}(H):= \braket{H^2} - \braket{H}^2$ is the variance of $H$ in the state $\ket\psi$.
Now one derives stationary conditions by setting the derivatives in $\dot\lambda_0$ and each $\dot\lambda_i$ to 0. The first condition gives
\begin{equation}
\label{eq:McLachlanGlobalPhase}
    \dot\lambda_0 = \braket{H} + \dot\lambda_i\imag\braket{\psi\vert\partial_i\psi}
    \,,
\end{equation}
while the remaining conditions in each $i$ are given by
\begin{equation}\dot\lambda_j\real\braket{\partial_i\psi\vert\partial_j\psi} + \dot\lambda_0\imag\braket{\psi\vert\partial_i\psi} - \imag\braket{\psi\vert H\vert\partial_i\psi} = 0
\end{equation}
Substituting for $\dot\lambda_0$, and defining the projection operators $Q_{\psi} := I - \ket{\psi}{\bra\psi} =: I - P_{\psi}$, one arrives at the final expression
\begin{equation}
    \left(\real\braket{\partial_i\psi\vert Q_\psi\vert\partial_j\psi}\right)\dot\lambda_j = \left(\imag\braket{\psi\vert H Q_\psi\vert\partial_i\psi} \right)\, .
\end{equation}
The matrix $M_{ij} := \real\braket{\partial_i\psi\vert Q_\psi\vert\partial_j\psi}$ and vector $V_{i} := \imag\braket{\psi\vert H Q_\psi\vert\psi}$ specify a linear system whose solutions give the McLachlan update vectors $\dot\lambda_i$ in each direction. 
Note that, while trivial, the global phase evolution can also be tracked via~\eqref{eq:McLachlanGlobalPhase}.

\bibliographystyle{utphys}
\bibliography{refs}

\end{document}